\documentclass[sn-mathphys]{sn-jnl}% Math and Physical Sciences Reference Style
\usepackage{hyperref}       % hyperlinks
\usepackage{url}            % simple URL typesetting
\usepackage{booktabs}       % professional-quality tables
\usepackage{amsfonts}       % blackboard math symbols
\usepackage{nicefrac}       % compact symbols for 1/2, etc.
\usepackage{microtype}      % microtypography
\usepackage{xcolor}         % colors

\jyear{2021}%

\theoremstyle{thmstyleone}%
\theoremstyle{thmstyletwo}%

\theoremstyle{thmstylethree}%

\graphicspath{{images/}}

\usepackage[normalem]{ulem}
\usepackage{color}
\begin{document}

\title[DRL for turbulent drag reduction in channel flows]{Deep reinforcement learning for turbulent drag reduction in channel flows}


\author*[1,2]{\fnm{Luca} \sur{Guastoni}}\email{guastoni@mech.kth.se}

\author[3]{\fnm{Jean} \sur{Rabault}}\email{jean.rblt@gmail.com}

\author[1,2]{\fnm{Philipp} \sur{Schlatter}}\email{pschlatt@mech.kth.se}

\author[4,2]{\fnm{Hossein} \sur{Azizpour}}\email{azizpour@kth.se}

\author*[1,2]{\fnm{Ricardo} \sur{Vinuesa}}\email{rvinuesa@mech.kth.se}

\affil*[1]{\orgdiv{FLOW, Engineering Mechanics}, \orgname{KTH Royal institute of Technology}, \orgaddress{\postcode{SE-100 44} \city{Stockholm}, \country{Sweden}}}

\affil[2]{\orgdiv{Swedish e-Science Research Centre (SeRC)}, \orgaddress{\postcode{SE-100 44} \city{Stockholm}, \country{Sweden}}}

\affil[3]{\orgdiv{IT Department}, \orgname{Norwegian Meteorological Institute}, \orgaddress{\street{Postboks 43}, \postcode{0313} \city{Oslo}, \country{Norway}}}

\affil*[4]{\orgdiv{School Elect. Eng. and Comp. Sci.}, \orgname{KTH Royal institute of Technology}, \orgaddress{\postcode{SE-100 44} \city{Stockholm}, \country{Sweden}}}


\abstract{We introduce a reinforcement learning (RL) environment to design and benchmark control strategies aimed at reducing drag in turbulent fluid flows enclosed in a channel.
The environment provides a framework for computationally-efficient, parallelized, high-fidelity fluid simulations, ready to interface with established RL agent programming interfaces. This allows for both testing existing deep reinforcement learning (DRL) algorithms against a challenging task, and advancing our knowledge of a complex, turbulent physical system that has been a major topic of research for over two centuries, and remains, even today, the subject of many unanswered questions. The control is applied in the form of blowing and suction at the wall, while the observable state is configurable, allowing to choose different variables such as velocity and pressure, in different locations of the domain.
Given the complex nonlinear nature of turbulent flows, the control strategies proposed so far in the literature are physically grounded, but too simple. DRL, by contrast, enables leveraging the high-dimensional data that can be sampled from flow simulations to design advanced control strategies.
In an effort to establish a benchmark for testing data-driven control strategies, we compare opposition control, a state-of-the-art turbulence-control strategy from the literature, and a commonly-used DRL algorithm, deep deterministic policy gradient. Our results show that DRL leads to 43\% and 30\% drag reduction in a minimal and a larger channel (at a friction Reynolds number of 180), respectively, outperforming the classical opposition control by around 20 and 10 percentage points, respectively.}




\maketitle
\section{Introduction}
Turbulent flows are ubiquitous both in nature and in engineering applications, from climate and weather dynamics~\cite{garratt1994atmospheric} to wind-turbine engineering~\cite{churchfield2012numerical}, and from turbulent blood streams in the human body~\cite{stein1976turbulent} to hypersonic flows around re-entry vehicles~\cite{schneider2004hypersonic}. Hence, turbulence is highly relevant both for economic and environmental reasons. Depending on the intended outcome, it can be desirable to promote turbulence, for example to enhance mixing in combustion engines~\cite{celik2001large}, or to hinder it, as a mean to reduce the drag on airplane wings and thus reducing the overall fuel consumption~\cite{spalart2011drag}. In both cases, some form of flow control needs to be designed and deployed. While flow control has attracted widespread attention over the years from different fields of the scientific community, turbulent flows exhibit a chaotic nature, they are multi-scale, highly non-linear and high-dimensional phenomena; furthermore, they are very expensive to simulate numerically in an accurate way. Still today, the challenges brought by turbulence prevent us from finding effective flow-control strategies in most realistic applications. 

Deep reinforcement learning (DRL) is a mathematical framework that has been used to design and learn control policies, also in physics research. This framework was successfully applied in optics~\citep{Nousiainen:21}, plasma physics \cite{degrave2022magnetic}, and thermodynamics~\citep{Beeler:19}. In fluid dynamics, the potential of DRL algorithms has been assessed for turbulence modelling~\cite{novati} and drag reduction~\cite{rabault}. 
Extending the latter application, in this work we introduce a RL environment to simulate a turbulent flow in a channel.
The setup chosen for the simulation is more computationally-efficient than other, more realistic geometries (\textit{e.g.} a wing or a turbine blade), while still being able to capture all the flow features in the vicinity of the wall. %[something about turbulence] Another important aspect is provided by the ..so the simulation should be performed using an optimized code.
The environment is based on the numerical solver SIMSON~\cite{chevalier}, which implements an efficient pseudo-spectral method to solve the Navier--Stokes equations for incompressible wall-bounded flows. The solver performs direct numerical simulations (DNSs), in which all the time and length scales are resolved without any approximation. This is essential to design a control strategy which does not exploit the limitations introduced by modelling some of the flow scales. Note that such a control policy would underperform if applied on a more realistic flow.  

This article is organized as follows: in section~\ref{s:related}, we review the most recent applications of deep learning and deep reinforcement learning in fluid mechanics, as well as the different approaches to reduce drag in channel flows from the literature. In section~\ref{s:meth}, we describe the simulation setup and how the numerical solver interacts with the deep reinforcement learning agent. The learning setup details are also provided. In section~\ref{s:exp} we validate our code against the opposition control results available in literature. Furthermore, the learning results in the minimal channel are reported, along with the drag reduction achieved by the DRL policy in two channel flow simulations of different sizes. Finally, in section~\ref{s:concl} we summarize our findings and outline some possible future developments. 
%[brief description of the state, actions and reward]

\section{Related work}\label{s:related}
\subsection{Deep learning in fluid mechanics}
Fluid-dynamics problems offer many opportunities and challenges for data-driven techniques. Recent research works published in machine-learning venues focused on improving the accuracy of coarse simulations~\cite{hoyer,zhang-learning} or approximating the dynamics of partial differential equations (PDEs)~\cite{anandkumar,wandel}. 
%Despite offering many opportunities and challenges for data-driven techniques, only few research works are occasionally published in machine-learning venues that target fluid-dynamics problems~\cite{otness,hoyer,anandkumar}.
At the same time, the application of deep-learning methods to turbulence requires specific domain knowledge, as testified by the large number of research works that have been published by domain specialists. % testifying the interest for deep learning methods.
A comprehensive review by Vinuesa and Brunton~\cite{enhancing} identifies three main areas of application of deep learning for fluid mechanics: the first possible application is to accelerate direct numerical simulations. A second area includes all the studies in which deep learning is used to enhance turbulence models in simulations, in order to make them less computationally expensive. Finally, neural-network architectures, such as autoencoders (AEs) can be used to develop reduced-order models of fluid flows~\cite{eivazi-ae}. 
%Reduce order models (autoencoders)?
On top of these applications, further applications have been envisioned~\cite{emerging}, including but not limited to temporal predictions of reduced-order models or spatial reconstructions of turbulent flows.
% Temporal predictions were first performed on low-dimensional dynamical systems that exhibit a chaotic behaviour. Two examples are the Kuramoto-Sivashinsky equation and the 9-equations model by Moehlis~\cite{moehlis_et_al}. 
% One [...]~\cite{eivazi}

Reconstruction of turbulent flow fields at a given instant has been attempted using convolutional networks~\citep{guastoni}. The possibility to perform \textit{non-intrusive sensing} of the flow, \textit{i.e.} sampling quantities without disrupting it, is an essential element to implement flow-control systems, which typically rely on velocity fields sampled at a given distance from the wall, as detailed in subsection~\ref{ss:dr}. %the subsection about drag reduction in channel flows later in this section.
Convolutional neural networks can also be trained to increase the resolution of coarse flow fields~\citep{fukami-super}. Note, however, that generative adversarial networks (GANs) have been proven to be more effective for this task~\citep{guemes-gans}.

\subsection{Reinforcement learning in fluid mechanics}
The application of (deep) reinforcement learning to fluid mechanics is still in its early phase compared with traditional supervised learning \cite{GARNIER2021104973, rabault2020deep}. Two main categories of contributions can be identified: the first focuses on modelling turbulence with the aid of DRL~\citep{novati}, while the second focuses on active-flow-control. In the following, we will focus on active flow control applications.

A number of contributions in this domain aim to control the movement of an agent in a fluid flow, for example, representing a fish swimming in a turbulent flow or in a fish school, where the reward aims to maximize the efficiency of the agent swimming~\citep{biferale,swimmers}. Another category of contributions use DRL to control the dynamics of the flow. The present work falls into this last subcategory. A notable example of such an application is provided by Ref.~\cite{rabault}, where the flow around a cylinder is controlled using jets orthogonal to the main flow direction. This case has been used as benchmark in a number of extension works~\cite{rabault2019accelerating, paris2021robust, tang2020robust, fan2020reinforcement, ren2021applying, xu2020active}, a fact that illustrates both the growing interest of DRL for active flow control, and the importance of providing benchmark cases that can be used as a starting point by the community. Recent works~\cite{varela} highlighted how different control strategies are selected by the DRL agent depending on the physical features of the flow to be controlled, and showed clearly that DRL agents can discover complex strategies not limited to simple opposition control. %hence it is fundamental to provide benchmarks that are physically-accurate.

Furthermore, DRL has been applied to a number of other control tasks, ranging from simple one-dimensional falling-fluid instabilities \cite{belus2019exploiting}, convection problems \cite{beintema2020controlling}, chaotic turbulent combustion systems \cite{bucci2019control} to a variety of engineering cases \cite{henry2021deep, Korb_2021, doi:10.1063/5.0052524, fluids7020062}.

In this study, we extend the application of DRL to active flow control in another category of flows, namely wall-bounded turbulent flows. This is a significant jump in complexity for at least two reasons. First, the two-dimensional (2D) cylinder case previously considered exhibits a well-defined shedding pattern at a specific frequency, which is dominant with respect to all the other dynamics, and suppressing it provides a straightforward drag-reduction strategy. If a higher Reynolds number is considered, a different policy is chosen: instead of reducing the shedding, the agent energizes the boundary layer on the cylinder surface to trigger the drag crisis~\cite{varela}. The channel flow, by contrast, is an inherently multi-scale phenomenon in which there are no obvious frequencies nor mechanisms that can be targeted in order to achieve drag reduction. Second, previous works focused on a 2D environment, while the channel flow is inherently 3D. This is a requisite for the development of true fully-featured turbulence structures, and it provides a more nonlinear, chaotic, and challenging benchmark to test novel DRL algorithms and control strategies, while also being closer to realistic full-scale configurations.
Very recent works have started exploring similar flow cases, for example in Ref.~\cite{sonoda} the use of DRL was tested in a standard channel flow (note that in this study we consider an open channel flow, as detailed in subsection~\ref{ss:sim}). Another notable example is Ref.~\cite{zeng}, where a Couette flow is controlled by means of two streamwise parallel slots. Note that in the latter case the training of the DRL agent is performed in a reduced-order model of the problem and then applied to the actual case. 

\subsection{Drag reduction in channel flows}\label{ss:dr}
Given the high importance of flow control in several fields, different techniques and frameworks have been presented in the literature. These control strategies are simple in nature, and exhibit a somewhat limited performance, but they are well established. We discuss these ´traditional´ control strategies in the following subsection, and we implement them in the solver to serve as a baseline.
In wall-bounded flows (such as our channel flow) the actuation is usually performed by means of a wall-normal velocity distribution applied at the wall, which corresponds to \textit{blowing} when the wall-normal velocity is positive, and \textit{suction} when it is negative. The control law is traditionally obtained by rescaling the wall-normal or spanwise velocity fluctuations at a given wall-normal sampling location $y_s$, and using it, with opposite sign, as the actuation value. The aim is to suppress the near-wall turbulent structures~\citep{opposition,opposition2,opposition3}. This strategy provides a drag reduction that depends on the height of the sensing plane, as detailed in subsection~\ref{ss:validation}. Similar drag reduction, at the corresponding Reynolds number, is reported for different flow cases, such as a turbulent boundary layer flow~\citep{comparison}.

\section{Methodology}\label{s:meth}

\subsection{Simulation setup}\label{ss:sim}
The geometry under consideration is an open channel flow. The wall is represented as a no-slip condition at the lower boundary, while a symmetry condition is imposed at the upper wall. Previous works~\citep{opposition,sonoda} used a standard channel flow for their investigations, with a no-slip condition applied to both the upper and lower boundary. We opted for an open channel flow because it allows for a better isolation of the flow features close to a solid boundary. In fact, in the full channel the large turbulent structures can extend beyond the channel centerline, with an effect also on the other wall. The dynamics of the near-wall turbulence close to the lower boundary is not affected by the boundary condition on the other side, providing a meaningful comparison between the two simulations and control techniques.

Throughout the paper, we use $(x,y,z)$ to indicate the streamwise, wall-normal and spanwise directions, respectively, and $(u,v,w)$ are used to indicate the corresponding velocity components.
We consider two different simulation domains, the first one is a \textit{minimal channel}~\citep{jimenez-minimal} with size $\Omega = L_x \times L_y \times L_z = 2.67h \times h \times 0.8h$ (where $h$ is the open-channel height). This domain size is large enough to simulate all the relevant near-wall statistical features of turbulence, while being computationally cheap. Note, however, that the limited spanwise dimension of the channel limits the simulation to a single low-speed streak, whereas in the larger domain, we are able to simulate the interaction of several of them. The second domain size is a larger channel of size $\Omega = L_x \times L_y \times L_z = 2\pi h \times h \times \pi h$. The friction Reynolds number is defined as $Re_{\tau} = u_{\tau}h/\nu$, where the friction velocity $u_{\tau}=\sqrt{\tau_w/\rho}$ (based on the the wall-shear stress $\tau_w$ and the fluid density $\rho$) and $\nu$ is the kinematic viscosity of the fluid. The Reynolds number represents the ratio between the inertial and viscous forces in the flow, determining the flow characteristics and how turbulent the flow is. We consider $Re_{\tau}=180$ in both domains, to compare the drag reduction with the corresponding results in Ref.~\citep{comparison}. 
The solver we used, SIMSON, is a pseudo-spectral code that uses the Chebyshev polynomials in the wall-normal direction. The resolution of the simulation is given by $N_x \times N_y \times N_z$, where $N_x$ and $N_z$ are the number of Fourier modes in the streamwise and spanwise directions, while $N_y$ is the number of Chebyshev modes in the wall-normal direction. The minimal channel is simulated with a resolution of $16 \times 65 \times 16$, while the resolution in the larger domain is $64 \times 65 \times 64$.
The time-advancement numerical scheme is a second-order Crank--Nicholson algorithm for the linear terms and a third-order Runge--Kutta method for the nonlinear terms.

\subsection{Reinforcement-learning environment and algorithm}
The simulation discussed above needs to be cast as a reinforcement-learning problem, defining the actions that can be performed, the state space and the reward function.

The control is performed by imposing a wall-normal velocity distribution. We cast the simulation as a multi-agent reinforcement-learning (MARL) problem, which implies that several independent agents cooperate in order to maximize the chosen reward function. In our setup all the agents operate locally and thus share the same actuation policy, which determines the action to perform based on the local observation of the environment provided to the agent. This approach has two important consequences: it allows us to avoid the curse of dimensionality on the DRL-control space dimension and to re-use the knowledge of the properties of the flow across the domain, as discussed in Ref.~\cite{belus2019exploiting}. 
We consider a grid of $N_{\rm{CTRLx}} \times N_{\rm{CTRLz}}$ agents that cover the entire lower wall of the channel. Here $N_{\rm{CTRLx}}$ and $N_{\rm{CTRLz}}$ represent the number of agents in the streamwise and spanwise directions, respectively. 
The policy learnt by the individual agent is translationally invariant to both the streamwise and spanwise directions.
Each agent computes the wall-normal velocity intensity to be applied at each evaluation.
The control policy is evaluated over a fixed time interval and the actual control varies linearly from the old value to the new one in order to avoid numerical instabilities related to sudden variation of the wall-normal velocity at the boundary. In our numerical experiments we use $\Delta t^+ = 0.6$ (where the time units are scaled with $t^* = \nu/u_{\tau}^2$).
The actuation value is limited to a prescribed range, in our case the agent can apply wall-normal velocities between $-u_{\tau}$ and $u_{\tau}$.

The agent does not have access to the entire velocity field, and the observed state consists of a portion of one or more wall-parallel planes of sampled flow quantities. The portion of the flow field that is observable by each agent is the one above the actuation. It is possible to sample any of the three velocity components at a given wall-normal location, the wall-normal location of the sampled plane is defined in the input file of the simulation and it is rounded up to the closest Chebyshev colocation node. 
For the remainder of the paper, we will refer to the different sampling heights using \textit{inner-scaled} wall-normal coordinates: $y^+ = y u_{\tau}/\nu = y/\ell^{*} \in [0,Re_{\tau}]$, where $\ell^{*}$ denotes the viscous length. %Each plane has size $N_x \times N_z$.

% The agent learns the scaling parameters $\mathbf{\alpha}$ for the velocity distribution at the sensing plane $y_s$ in the expression
% \begin{equation}
%     v(x,0,z,t) = -\mathbf{\alpha}(z,t)\big(v(x,y_s,z,t)- \langle v(x,y_s,z,t) \rangle \big).    
% \end{equation}
% Here $\langle v(x,y_s,z,t) \rangle$ is the spatial mean of the wall-normal velocity field that ensures that the control has a zero-net-mass-flux within the domain. $\mathbf{\alpha}$ is vector of positive scaling parameters. The number of parameters $n_{\mathrm{CTRL,z}}$ is set in the simulation input file. The domain is divided in $n_{\mathrm{CTRL,z}}$ equal slices in the spanwise direction and in each slice the scaling parameter is constant.
% The parameters can vary in a prescribed range, in particular we set $|\alpha_i| \in [0,1]$.

The same global reward value is provided to all MARL agents, defined as the percentual reduction of the wall-shear stress with respect to the uncontrolled flow, averaged over the entire wall. We do not consider the instantaneous value of this quantity, rather an averaged value between two actions is returned. The solver performs multiple time-step advancements between one action and the next one because the time step required for the stability of the numerical method is much lower than the actuation time. The actuation interval is also a parameter that can be defined in the input file and it should be set to be small enough to allow it to react to the change of the small flow scales, and sufficiently large to observe the effect of the action at the intermediate scales~\cite{belus2019exploiting}. Since turbulence is a multi-frequency phenomenon, the reward signal reflects this characteristic and the reward averaging helps to provide a more reliable estimate of the actual wall-shear-stress reduction than an instantaneous value would do.
Note that the presence of turbulence in the flow induces a higher drag with respect to a laminar flow and the implemented control effectively reduces the level of turbulence. In fact, the control can theoretically induce a re-laminarization of the flow. This condition represents an upper bound for the drag reduction achievable with the control~\citep{FUKAGATA20091082}. %check this ref
This value depends on the chosen Reynolds number and in our case is 73.9\%. Note, however, that there is no guarantee that this value is ever attainable with any control law.

The deep reinforcement learning algorithm that is used to train the agent is the deep deterministic policy gradient (DDPG)~\cite{ddpg}. As the name suggests, this algorithm optimizes a deep neural network that approximates the relation between the state and the action to be performed (\textit{i.e.} a policy function). After learning, the policy is deterministic. During learning, Gaussian noise with zero mean and variance $0.1u_{\tau}$ is added to foster exploration.
The policy is updated every $\Delta t^+ = 180$, with 64 mini-batch gradient updates. The mini-batches are sampled from a buffer replay that includes 5,000,000 (state, action, reward) tuples.  

\subsection{Solver-DRL interface}
The reinforcement-learning framework chosen for this work is Stable-baselines~\citep{stable}, while the environment is coded as a custom PettingZoo/Gym environment~\cite{pettingzoo}. Both Stable-baselines and the Gym environment are written in Python, however the fluid solver is coded in FORTRAN 77/90. An interface between the two programming languages is then needed in order to communicate the quantities (state, control values, reward) that are necessary for the learning to take place.
While previous studies have coupled the solver and the RL algorithm using an input/output (I/O) stream~\cite{wang}, in this case the interface is based on message-passing interface (MPI). The fluid solver is spawned as a child process of the main learning process and an intercomm is created between the solver and the Python main program. 
After the initialization, the solver waits for information requests from the main program. Once the request is received by the solver, further MPI-based messages are exchanged between the agent and the solver, depending on the requested information.
Communication requests are handled using five-character strings, which determine the sequence of instructions that the solver has to perform for each interaction of the agent with the environment. The admissible values for the request string are:
\begin{itemize}
    \item \texttt{STATE}: this string is used to request the simulation to communicate the state to the agent.%the current value of the state is communicated to the agent.
    \item \texttt{CNTRL}: after this request, the agent communicates the new action to the environment, in particular the new values of the controllable parameters are passed to the fluid solver. 
    \item \texttt{EVOLVE}: once the control parameters are updated with \texttt{CNTRL}, several time iterations of the solver are computed in order to observe the effect of the chosen action on the environment. The instantaneous wall-shear stress is passed to the Python program and it is used to compute the reward. %stored in a ring buffer. The final reward is obtained after a given number of iterations by averaging the data stored in the ring buffer.
    \item \texttt{TERMN}: this request interrupts the solver and closes the FORTRAN program. This request is used at the end of each episode before restarting the environment. The initial condition is the same for every episode unless stated otherwise.
\end{itemize}

The combination of these messages is used to define the required functions for the Gym environment application programming interface (API).

\section{Experiments}\label{s:exp}
\subsection{Validation and baseline}\label{ss:validation}
As mentioned in section~\ref{s:related}, opposition control is a simple established control benchmark that is well understood theoretically and can be used as a reference to assess DRL control laws. Given the velocity distribution at the sensing plane $y_s$, the actuation at the wall $v_\mathrm{w}$ applied by opposition control can be computed with the expression:
\begin{equation}
    v_\mathrm{w}(x,z,t) = -\alpha\big[v(x,y_s,z,t)- \langle v(x,y_s,z,t) \rangle \big].    
\end{equation}
Here $\langle v(x,y_s,z,t) \rangle$ is the spatial mean of the wall-normal velocity field, and the second term of the equation enforces zero-net-mass-flux within the domain, so that the actuation plane as a whole is not adding or removing fluid mass to the flow. $\alpha$ is a positive scaling parameter which is fixed in the spatial directions and time and its typical value is $\alpha=1$, independently from the height of the sampled velocity plane.
In order to validate our code implementation we reproduced this control strategy in our solver, letting the environment evolve with constant $\alpha$ for a sufficiently long time, such that the drag-reduction rate has reached a stationary value.
In Ref.~\cite{opposition}, the highest drag reduction was found by sampling the wall-normal velocity at $y^+=10$. The computational domain had size $\Omega = L_x \times L_y \times L_z = 4\pi h \times h \times 4/3\pi h$, with resolution $N_x \times N_y \times N_z = 128 \times 129 \times 128$. The friction Reynolds number $Re_{\tau}$ was reported to be 180 and the Reynolds number based on the centerline velocity $U_{cl}$ (the one at the top boundary, in the case of the open channel) is in close agreement with our setup: the reported value was $Re_{cl}=3300$, while our simulation is performed with $Re_{cl}=3273$. 
The drag reduction reported in Ref.~\cite{opposition} was $\approx14\%$ when sampling at $y^+=10$. Our result in the larger channel is 17.73\%. This is in acceptable agreement with that of the literature and further validates our simulation, considering the lower resolution and taking into account the small difference in spanwise size of the domain. 

% the friction Reynolds number $Re_{\tau}$ is not reported, instead they use the Reynolds centerline number, which is based on the centerline velocity $U_{cl}$ (the one at the top boundary, in the case of the open channel). The reported value is $Re_{cl}=1800$, while ours simulation is performed with $Re_{cl}=4200$. The reported drag reduction is $\approx25\%$ when sampling at $y^+=10$. Our result is $\approx19\%$, possibly because of the higher $Re_{cl}$.
A different sampling plane and slightly lower $Re_{cl}$ were used in Ref.~\citep{opposition2}. The domain here is $\Omega = L_x \times L_y \times L_z = 4\pi h \times h \times 2\pi h$, with resolution $N_x \times N_y \times N_z = 256 \times 130 \times 256$. When rescaling the velocity at $y^+=15$ with $Re_{cl}=3240$, the resulting drag reduction was $\approx25\%$ and our own test resulted in a reduction of $25.67\%$, which is in very good agreement. %, even if our Reynolds number is slightly higher. %slightly lower, but our Reynolds number is slightly higher.
Finally, in Ref.~\cite{comparison}, the field at $y^+=15$ at $Re_{\tau}=180$ is sampled. In this case, the resolution is slightly lower than those of the previous studies the wall-parallel directions ($N_x \times N_y \times N_z = 160 \times 257 \times 128$). On the other hand, the channel size is also lower, being exactly the same as our larger channel. They reported $\approx24\%$ drag reduction, and our result is also in good agreement with theirs. 
Overall, the results from our implementation match the ones reported in the literature, giving us confidence in our simulation setup and control implementation. The drag reduction obtained with these settings is used as baseline for comparison with the DRL results. 

\subsection{Minimal channel learning and testing}\label{ss:MClearn}
\begin{figure}
\begin{center}
\includegraphics[width=.47\textwidth]{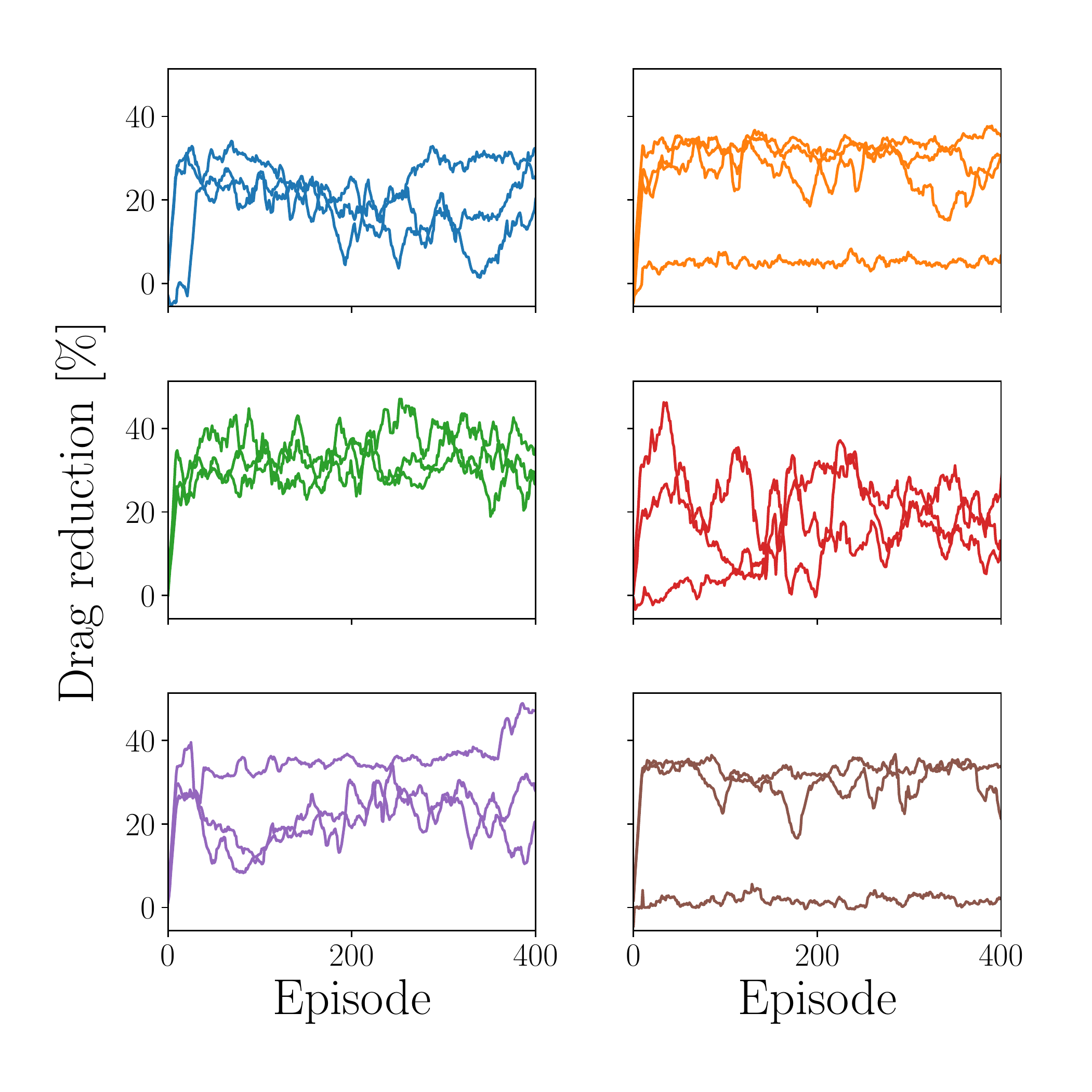}
\includegraphics[width=.47\textwidth]{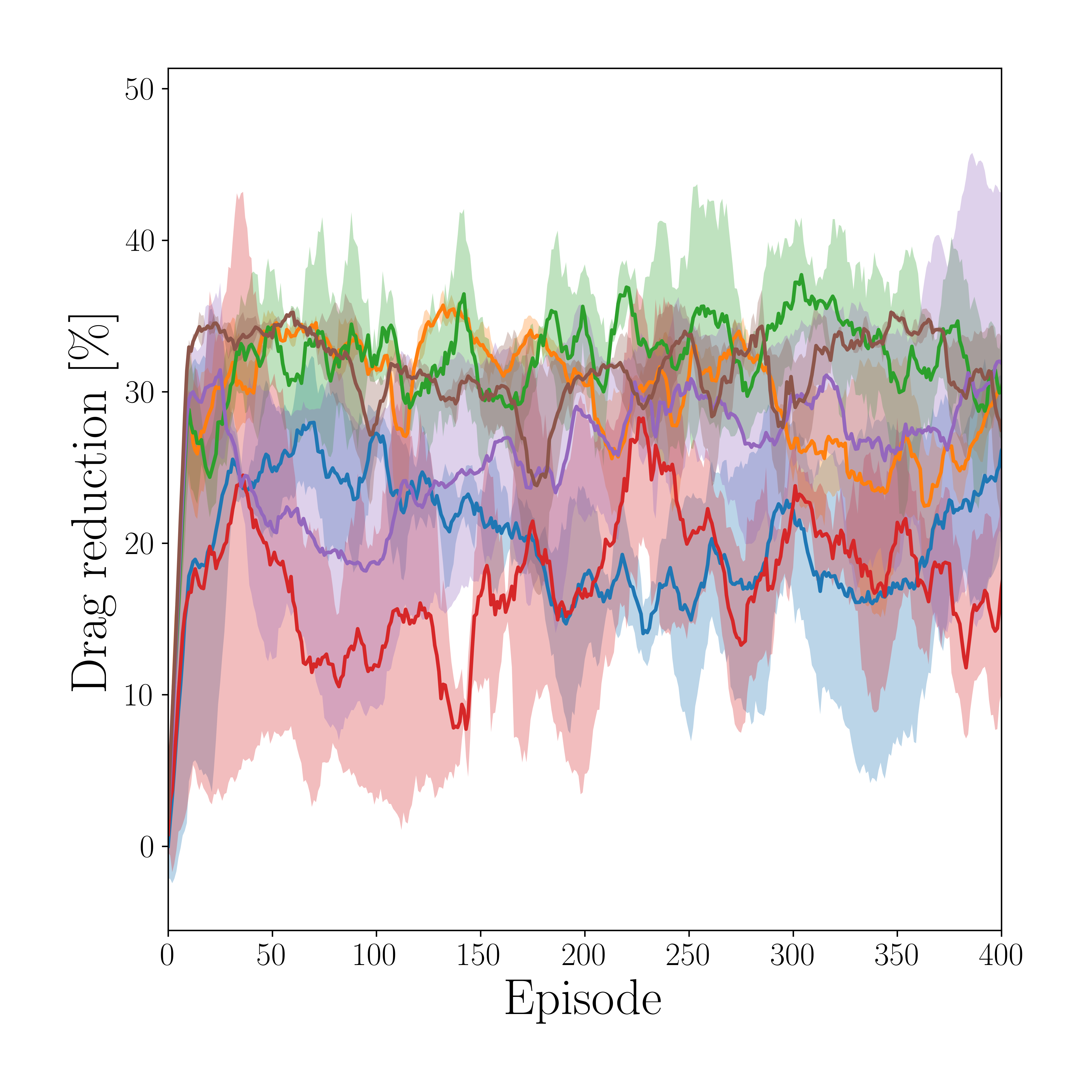}
\end{center}
\caption{\label{fig:MC-learn} (Left) Running mean of the drag reduction with respect to the reference uncontrolled case during agent learning in the minimal channel. Each panel shows 3 different learning runs for 6 initial conditions. (Right) Running mean of the drag reduction with respect to the reference uncontrolled case during learning in the minimal channel, averaged over different learning runs. The shaded area represents the variance of the drag reduction with the different runs.}
\end{figure}

We first consider the smaller domain, {\it i.e.} the minimal-channel configuration. In this case, the number of agents is set to $N_{\rm{CTRLx}} = N_x$ and $N_{\rm{CTRLz}} = N_z$.
The learning is performed using several initial conditions for 400 episodes. The streamwise and wall-normal components of the velocity at $y^+=15$ are provided to the agents as the observable state. In each learning, all the episodes are initialized with the same flow field. Figure~\ref{fig:MC-learn} highlights the sensitivity of the learning to the initial condition of the episode. One important remark is that (D)RL algorithms are based on a trial-and-error approach to the reward optimization. This means that not all learning runs provide a drag-reducing policy, as shown in the aforementioned figure. Note that the unsuccessful runs (shown in the left panel) are removed from the average for clarity. 

%shows that on average we are able to achieve about 20\% drag reduction. On the other hand, the learning is found to be very sensitive to the initial condition. Depending on the initialization, it is possible to achieve a very good performance or no learning at all. If the training is performed with a random initial condition at each episode, the learning progress is qualitatively similar to the average of the individual runs with a fixed initial condition. 

\begin{figure}
\begin{center}
\includegraphics[width=.47\textwidth]{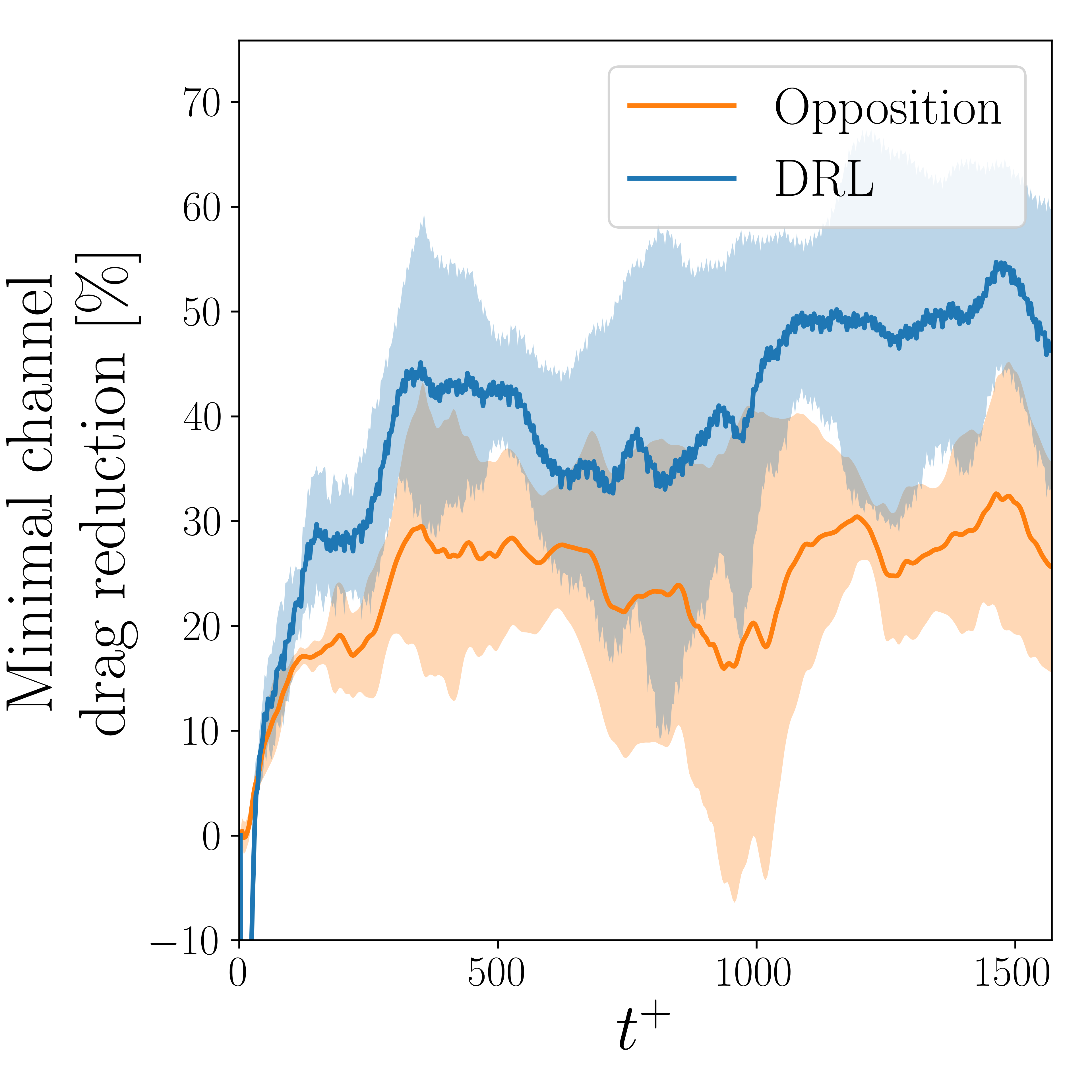}
\includegraphics[width=.47\textwidth]{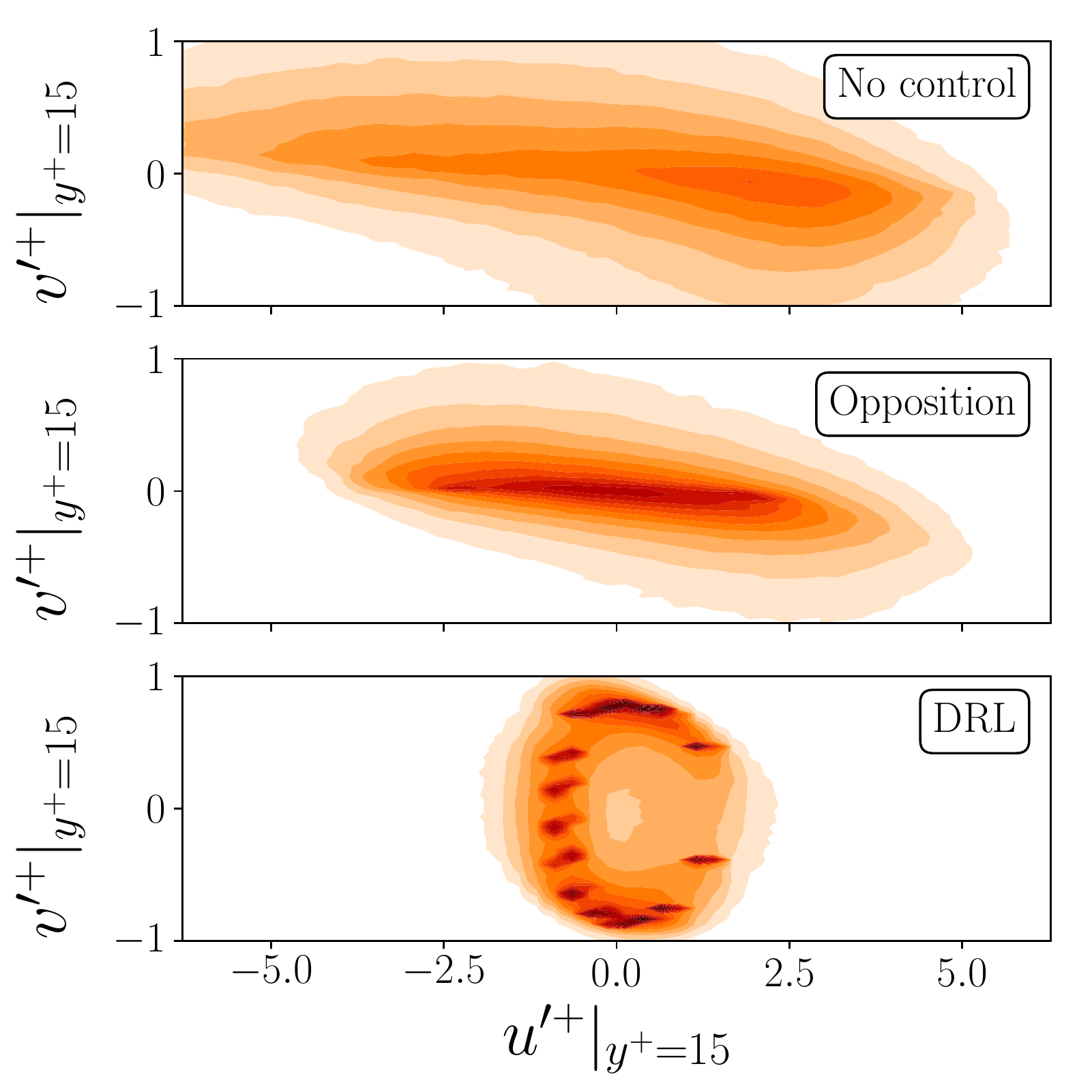}
\end{center}
\caption{\label{fig:MC-test} (Left) Drag reduction with respect to the uncontrolled case obtained in the minimal channel, when using the policy learnt using DRL or opposition control. The result is averaged over 6 different initial conditions. The shaded area represents the variance of the drag reduction with the different initial conditions. (Right) Distribution of the inner-scaled velocity-fluctuation components after the initial transient ($t^+>500$) in the streamwise ($u$) and wall-normal ($v$) directions, for the uncontrolled case (top), with opposition control (middle) and when using DRL (bottom).}
\end{figure}

Even though the learning has a strong sensitivity on the initialization of the episode, the learnt policies perform consistently during testing, regardless of the selected initial condition. Figure~\ref{fig:MC-test} shows the drag reduction achieved with one of the best-performing policy. The policy test is repeated for six different initial conditions, in order to assess the generalization of the learnt control policy. % learnt with the most favourable initial condition, tested on all the learning initialization. 
The performance is compared with the baseline strategy (opposition control, with sensing plane at $y^+=15$) for the same set of initial conditions. Initially, the DRL policy produces a strong increase in the drag for a brief time, corresponding to a negative value for the drag reduction. After this drag increase, it is possible to observe how the average drag reduction after the initial transient is consistently higher than the one obtained with opposition control: the DDPG policy provides 43\% drag reduction, while opposition control is limited to 26\%, as mentioned in subsection~\ref{ss:validation}.

Figure~\ref{fig:MC-test} (right) shows the effect of the control on the velocity-fluctuations distribution. Using the friction velocity for the uncontrolled case, it is possible to observe how the two control approaches produce different changes in the fluctuations distribution. With opposition control, the range of the fluctuations is reduced, but the overall shape of the distribution is unchanged. In fact, from the perspective of quadrant analysis~\cite{wallace_eckelmann_brodkey_1972,lu_willmarth_1973}, sweeps and ejections remain the dominant features in the near-wall region. On the other hand, DRL has a more significant effect on the distribution: at the sensing plane, we can observe a distribution that is almost symmetric with respect to the wall-normal fluctuations. With the DRL control, the range of the streamwise fluctuations is greatly reduced, while the wall-normal fluctuations are increased. Furthermore, the predominance of sweeps and ejections vanishes, with a more even distribution of events among the four quadrants. Consequently, DRL learns a control strategy with a profound impact on the flow physics. Analyzing the sensing plane at $y^+=15$ when the control is applied reveals that the wall streak in the minimal channel is significantly attenuated by a streamwise travelling wave.

% \subsection{\revd{Larger channel learning}}\label{ss:LClearn}
% \revd{More numerical experiments are performed in a larger domain. Also in this case, we considered $N_{\rm{CTRLx}} = N_x$ and $N_{\rm{CTRLz}} = N_z$. Note, however, that this implies that a larger number of agents cooperates within a single instance of the simulation. This results in a larger number of (state, action, reward) tuples that are experienced at each interaction with the environment.}

\subsection{Larger channel testing}\label{ss:LCtest}
\begin{figure}
\begin{center}
\includegraphics[width=.47\textwidth]{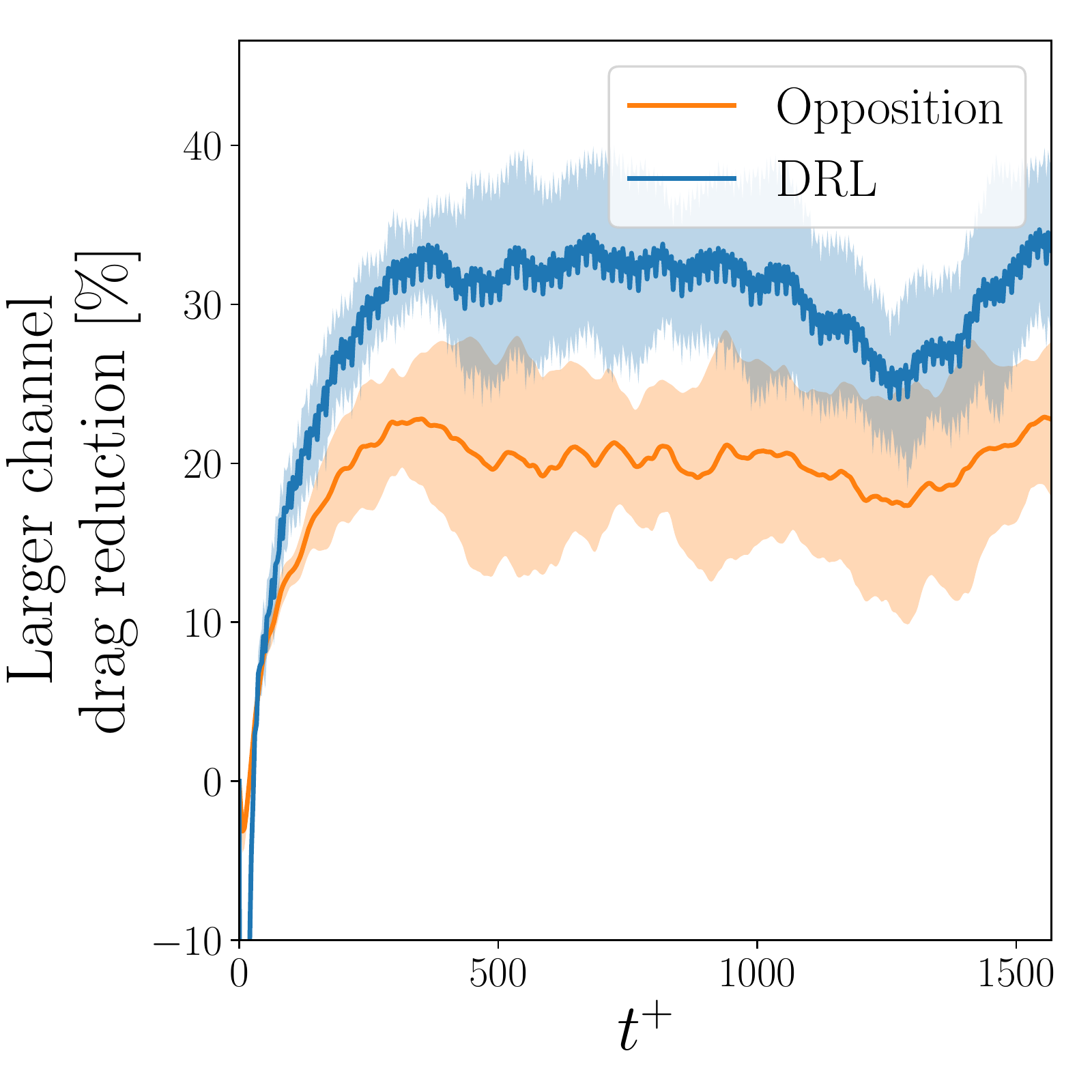}
\includegraphics[width=.47\textwidth]{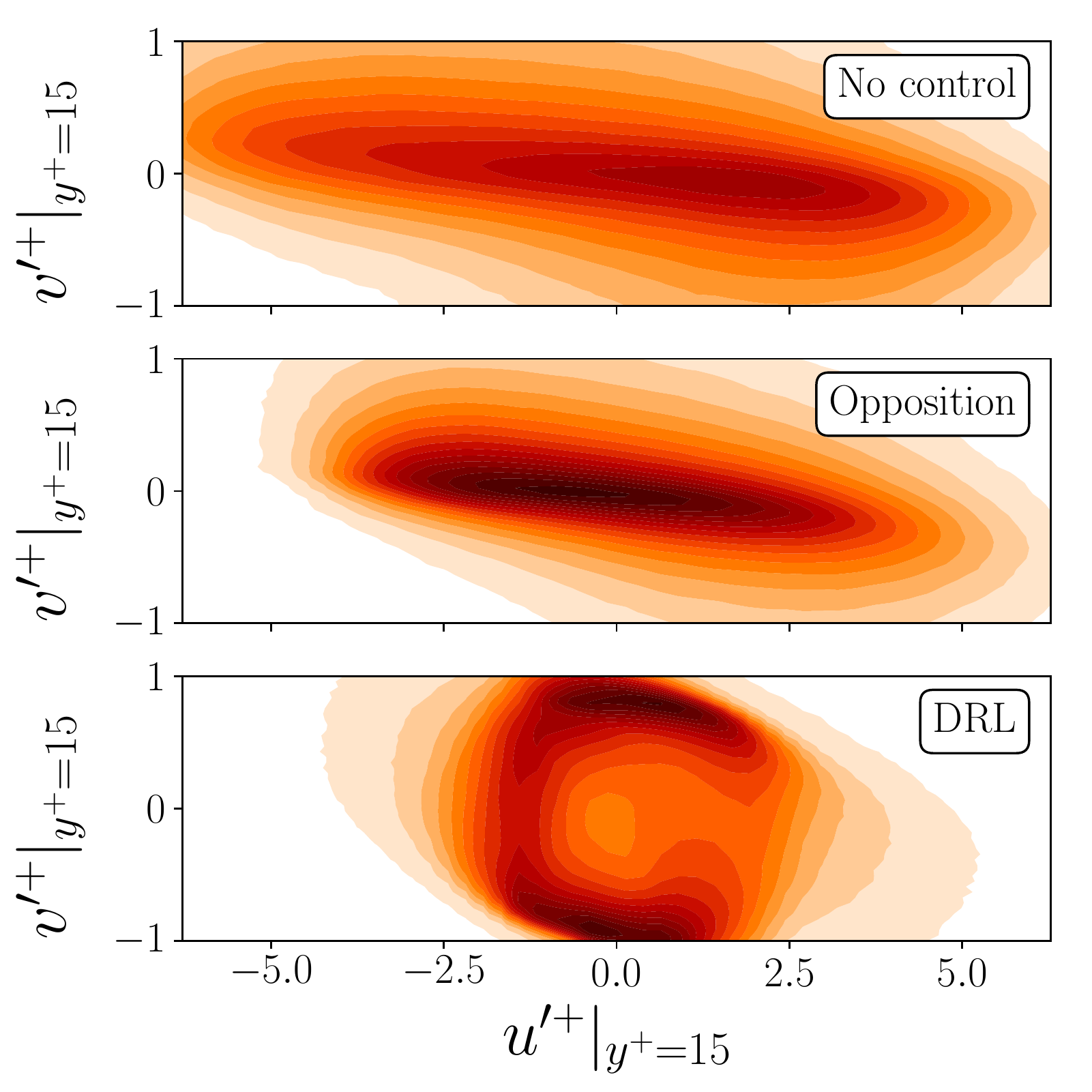}
\end{center}
\caption{\label{fig:LC-test} (Left) Drag reduction with respect to the uncontrolled case obtained in the larger channel, when using the DRL policy learnt in the minimal channel versus using opposition control. The result is averaged over 6 different initial conditions. The shaded area represents the variance of the drag reduction with the different initial conditions. (Right) Distribution of the inner-scaled velocity-fluctuation components after the initial transient ($t^+>500$) in the streamwise ($u$) and wall-normal ($v$) directions, for the uncontrolled case (top), with opposition control (middle) and when using DRL (bottom).}
\end{figure}

One of the relevant features of the learnt policy is that it is local and translational invariant, meaning that it can be applied with no modification to different domain sizes and flow cases, provided that the size of the state observations and of the reward are the same. The most important assumption behind the application of the same policy is that the underlying physical features exploited by the agent in the environment where the learning is performed are also present in the new environment.
When applying the policy to a larger domain, we still considered $N_{\rm{CTRLx}} = N_x$ and $N_{\rm{CTRLz}} = N_z$. Note, however, that this implies that a larger number of agents cooperate within a single instance of the simulation. %This results in a larger number of (state, action, reward) tuples that are experienced at each interaction with the environment.
The left panel of figure~\ref{fig:LC-test} shows the drag reduction achieved in the larger domain when the policy learnt in the minimal channel is applied. Remarkably, the policy provides a drag reduction that is higher than the opposition-control baseline, regardsless of the flow initial condition. The effectiveness of the control policy is tested without any further tuning in the larger domain. The success of this transfer application shows that the DRL agent has been able to learn a control strategy that robustly exploits the features in the flow, so that the learnt policy can be effective in different cases which have a similar physical characterization. %on a longer time horizon with respect to the one used for the learning episodes. 
The magnitude of the reduction in this case is smaller than the one obtained in the minimal channel, with the DDPG policy providing 30\% drag reduction, still performing better than opposition control, which yields 20\%. 
One possible explanation for this result is the fact that the policy is learnt in the minimal channel, where some physical features, such as the interaction of the streaks, cannot be experienced by the agent. Further evidence of this difference can be found by assessing the effect of the two control strategies on the velocity-fluctuations distribution shown in the right panel of figure~\ref{fig:LC-test}. For the DRL control, the symmetry with respect to the wall-normal fluctuations is more pronounced than before, showing a high probability of strong wall-normal fluctuations, coupled with weak streamwise fluctuations. In this case the predominance of sweeps and ejections is also reduced by the DRL in the larger domain, but it is not eliminated to the extent that is observed in the minimal channel. 

\section{Conclusions}\label{s:concl}
In this paper we have described the implementation of a pseudo-spectral solver for DNS of fluid flows as a reinforcement-learning environment, with which the agents interact to maximize the drag reduction within the domain. The environment supports three-dimensional, fully-turbulent flow simulations, allowing for the discovery of physically-accurate control policies. The environment can be customized with different flow cases and adapted to tasks of varying difficulty, selecting a different number of input quantities or a different number of agents, for instance. 
Using the environment, we applied DRL for drag reduction in an open channel flow, with two different domain sizes and resolutions. The DDPG algorithm allows the identification of control policies that on average show a higher drag reduction than the one provided by opposition control, used here as baseline. In the minimal channel, the control strategy yields 43\% drag reduction, while in the larger domain the achieved reduction is smaller (30\%), but still consistently better than the baseline. The performance improvement with respect to the baseline is around 20 percentage points in the minimal channel and 10 percentage points in the larger domain.
None of the policies learnt or used as baseline considers the energetic cost of the actuation, meaning that the net-energy saving is lower than the figures provided here. In this regard, reinforcement learning could in further work help designing more efficient control strategies by incorporating explicitly the energy cost of the actuation as part of the reward function.
Currently, the learnt policies are not able to re-laminarize the flow. Since this condition represents the upper bound for drag reduction, an improvement of the control policy is still theoretically possible, although there is no guarantee that this is attainable through DRL control or other techniques. %\rev{[LG] higher actuation could work but maybe this is a bit speculative}
The current study focuses on a open-channel flow with uncontrolled friction Reynolds number $Re_\tau = 180$. Our setup can also be adapted to different flow cases, paving the way to applying similar techniques to increasingly complex systems such as boundary layers or higher Reynolds numbers.
Simulating increasingly high Reynolds numbers becomes progressively more computationally expensive, however the simulation wall-clock time can be reduced by using the MPI parallelization of the solver and HPC clusters. 
The possibility to change the Reynolds number of the flow simulation represents a way to ´tune´ the difficulty of the control problem, making the drag reduction in a open channel flow an ideal benchmark to test new policies.
Increasing the Reynolds number is necessary step in order to design a control strategy that can be applied in practical applications; on the other hand, it also represents an appealing research direction as the maximum achievable drag reduction increases with the Reynolds number. 
It must be noted that a control policy that yields drag reduction at a given Reynolds number, may not be as effective at a higher $Re$ because different physical mechanisms can be responsible for the drag~\citep{marusic}. In this regard, deep reinforcement learning constitutes a promising framework, thanks to its end-to-end approach: all the physical features of the turbulent flow are represented in the environment, allowing for new and possibly more effective control strategies.

\backmatter

\bmhead{Acknowledgments}
The authors acknowledge the Swedish National Infrastructure for Computing (SNIC) for providing the computational resources by PDC, used to carry out the numerical simulations. The authors acknowledge Prof. Yosuke Hasegawa for insightful discussions on the DRL-algorithm setup.

\section*{Declarations}

\begin{itemize}
    \item \textbf{Funding} This work is supported by the founding provided by the Swedish e-Science Research Centre (SeRC), ERC grant no.~"2021-CoG-101043998, DEEPCONTROL" and the Knut and Alice Wallenberg (KAW) Foundation.
    \item \textbf{Conflict of interest} The authors report no conflicts of interest
    \item \textbf{Code availability} The datasets generated during the current study as well as all the codes are included in the repository \url{https://github.com/KTH-FlowAI/MARL-drag-reduction-in-wall-bounded-flows}.%will be made available open access as soon as the article is published.
    \item \textbf{Authors' contributions} The general idea of the project was developed by Ricardo Vinuesa (RV). The deep reinforcement learning (DRL) interface for the flow simulation was developed by Luca Guastoni (LG), with input by RV, Philipp Schlatter (PS) and Jean Rabault (JR). The learning runs and post-processing of the results were performed by LG. JR, RV and Hossein Azizpour (HA) provided feedback. The paper was written by LG and edited by RV and JR. Comments were provided by RV, JR, HA and PS.
\end{itemize}

\begin{appendices}

\section{Effect of the episode length}\label{secA1}

\begin{figure}
\begin{center}
\includegraphics[width=.47\textwidth]{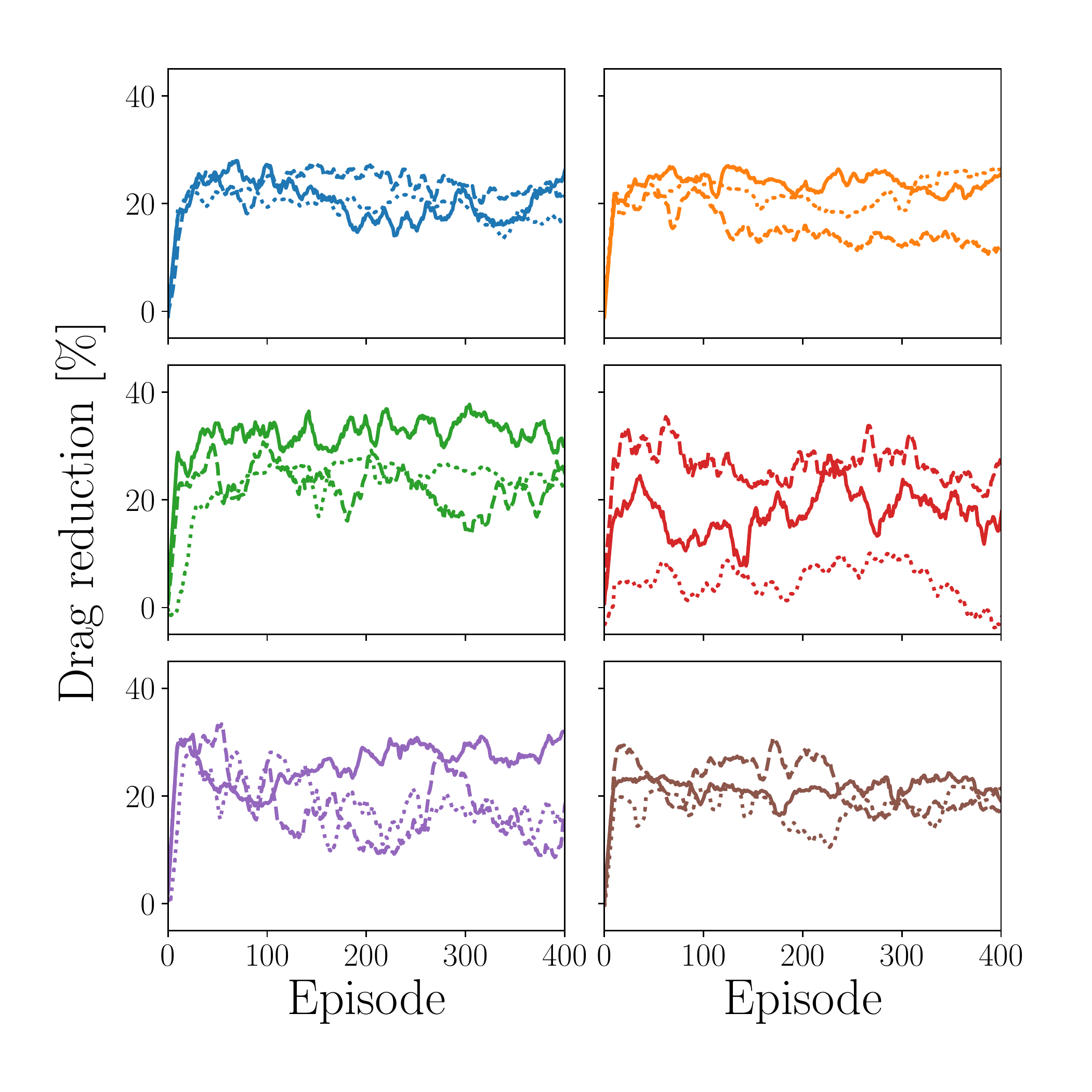}
\end{center}
\caption{\label{fig:MC-eplength} Running mean of the drag reduction with respect to the reference uncontrolled case during learning in the minimal channel, averaged over different learning runs. Each panel shows 3 different learning runs for 6 initial conditions.}
\end{figure}

In subsection \ref{ss:MClearn}, we consider episodes of 3000 interactions between the environment and the agent. This corresponds to a simulation time of $t^+ \approx 1500$. Here we verify the effect of the episode length on the training by reducing the number of interactions to 1000 and 2000, as shown in figure \ref{fig:MC-eplength}. The initial choice of the episode length is designed to include the initial transient (typically $t^+ \approx 500$) and a sufficiently long time after that. However, in our experiments, longer episodes do not provide a significant advantage in terms of drag reduction. It is possible to observe how the learning appears to improve faster at the very beginning of the learning but this is simply related to the higher number of interactions per episode. Shorter episode might provide a way to reduce the overall time and computational cost of the learning, which increase quickly with the number of agents and the resolution/Reynolds number, respectively.

\end{appendices}

\bibliography{bibliography}

\end{document}